\documentstyle[12pt,epsfig]{article}

 \catcode`\@=11
\def\Bbb{\ifmmode\let\next\Bbb@\else
 \def\next{\errmessage{Use \string\Bbb\space only in math mode}}\fi\next}
\def\Bbb@#1{{\Bbb@@{#1}}}
\def\Bbb@@#1{\fam\msbfam#1}

\if@twoside
\else
\fi
\ifcase \@ptsize
\or 
\or 
\fi

\def\@citex[#1]#2{%
\if@filesw \immediate \write \@auxout {\string \citation {#2}}\fi
\@tempcntb\m@ne \let\@h@ld\relax \def\@citea{}%
\@cite{%
  \@for \@citeb:=#2\do {%
    \@ifundefined {b@\@citeb}%
      {\@h@ld\@citea\@tempcntb\m@ne{\bf ?}%
      \@warning {Citation `\@citeb ' on page \thepage \space undefined}}%
      {\@tempcnta\@tempcntb \advance\@tempcnta\@ne%
      \@tempcntb\number\csname b@\@citeb \endcsname \relax%
      \ifnum\@tempcnta=\@tempcntb 
        \ifx\@h@ld\relax%
          \edef \@h@ld{\@citea\csname b@\@citeb\endcsname}%
        \else%
          \edef\@h@ld{\ifmmode{-}\else--\fi\csname b@\@citeb\endcsname}%
        \fi%
      \else
        \@h@ld\@citea\csname b@\@citeb \endcsname%
        \let\@h@ld\relax%
      \fi}%
    \def\@citea{,\penalty\@highpenalty\,}%
  }\@h@ld
}{#1}}
\@addtoreset{equation}{section}
\def\section{\@startsection {section}{1}{\z@}{-3.5ex plus -1ex minus
 -.2ex}{2.3ex plus .2ex}{\large\bf\centering}}
\def\subsection{\@startsection{subsection}{2}{\z@}{-3.25ex plus -1ex minus
 -.2ex}{1.5ex plus .2ex}{\sc}}
\gdef\@publabel{\hfil}
\gdef\@pubdate{\null}
\gdef\@pubnumber{\null}
\gdef\@author{\null}
\gdef\@title{\null}
\gdef\@abstract{\null}
\long\def\pubdate#1{\gdef\@pubdate{#1}}
\long\def\pubnumber#1{\gdef\@pubnumber{#1}}
\long\def\publabel#1{\gdef\@publabel{#1}}
\long\def\author#1{\gdef\@author{#1}}
\long\def\title#1{\gdef\@title{#1}}
\long\def\abstract#1{\gdef\@abstract{#1}}
\def\titlerelax{
\let\maketitle\relax
\let\settitleparameters\relax
\let\consolidatetitle\relax
\let\inittitlepage\relax
\let\finishtitlepage\relax
\let\titlepagecontents\relax
\let\multithanks\relax
\let\titlebaselines\relax
\let\@makepub\relax
\let\@maketitle\relax
\let\@makeauthor\relax
\let\@makeabstract\relax
\let\@maketitlenote\relax
\let\thanks\relax
\let\titlerelax\relax}
\def\titleclean
{\gdef\@titlenote{}
\gdef\@abstract{}
\gdef\@author{}
\gdef\@title{}
\gdef\@pubdate{}\gdef\@pubnumber{}\gdef\@publabel{}
\gdef\@dpublabel{}
}

\def\@makepub{\vbox to \z@{\hbox to \textwidth{\hfill
\@publabel \hfill
\llap{\parbox[t]{0.25\textwidth}{\raggedleft\@pubnumber}}}%
\vss}}
\def\@maketitle{\vskip 60pt \begin{center}
 {\LARGE \@title \par}
 \end{center}}
\def\@makeauthor{{%
\def\and{\smallskip {\normalsize \rm and\smallskip }}
\def\And{\medskip {\normalsize \rm and\\}\medskip}
\long\def\address##1{{\def\and{\\and\\}\medskip
				{\small \it \\##1\\}
}}
{\centering
 \vskip 3em
 \large \lineskip .75em
 \@author}
 \par}}
\def\@makedate{\vskip 1.5em
 {\raggedright \small \noindent\@pubdate \par}}
\def\@makeabstract{\vskip 1.5em
{\small
\begin{center}
{\bf ABSTRACT\vspace{-.5em}\vspace{0pt}}
\end{center}
\quotation \@abstract \endquotation}}
\def\maketitle{\titlepage
\let\footnotesize\small \setcounter{page}{0}
\@makepub
\vfil
\@maketitle
\@makeauthor
\vfil
\@makeabstract
\@thanks
\vfil
\@makedate
\if@restonecol\twocolumn \else \eject \fi
\titlerelax \titleclean
\setcounter{footnote}{0}
}

\catcode`\@=12

\begin{document}
\bibliographystyle{npb}

\def\be{\begin{equation}}
\def\ee{\end{equation}}
\let\b=\beta
\def\blank#1{}

\def\cdd{{\cdot}}
\def\cev#1{\langle #1 \vert}
\def\cH{{\cal H}}
\def\comm#1#2{\bigl [ #1 , #2 \bigl ] }
\def\compact{ reductive}
\def\cont{\nonumber\\*&&\mbox{}}
\def\cO{{\cal O}}
\def\cul #1,#2,#3,#4,#5,#6.{\left\{ \matrix{#1&#2&#3\cr #4&#5&#6} \right\}}

\def\dz{Dz}
\def\dz{\hbox{$d\kern-1.1ex{\raise 3.5pt\hbox{$-$}}\!\!z$}}
\def\dz{ \frac{d\!z}{2\pi i}}
\def\en{\end{equation}}
\def\enn{\end{eqnarray}}
\def\eq{\begin{equation}}
\def\eqq{\begin{eqnarray}}



\def\half#1{\frac {#1}{2}}

\def\ip#1#2{\langle #1,#2\rangle}


\def\k{k}


\def\Mf#1{{M{}^{{}_{#1}}}}
\def\mno{{\textstyle {\circ\atop\circ}}}
\def\mod#1{\vert #1 \vert}

\def\Nf#1{{N{}^{{}_{#1}}}}
\def\ni{\noindent}
\def\no{{\textstyle {\times\atop\times}}}
\def\no:#1:{\mno#1\mno}
\def\nox{{\scriptstyle{\times \atop \times}}}


\let\p=\phi
\def\posdef{ positive-definite}
\def\posdefness{ positive-definiteness}
\def\Qf#1{{Q{}^{{}_{#1}}}}
\def\Qstar{\mathop{\no:QQ:}\nolimits}

\def\reductive#1#2{#1}


\def\tr{\mathop{\rm tr}\nolimits}
\def\Tr{\mathop{\rm Tr}\nolimits}







\def\vec#1{\vert #1 \rangle}
\def\vac{\vec 0}

\def\wan{$\WA_n$ }
\def\Wb{\bar W}
\def\Wf#1{{W{}^{{}_{#1}}}}
\def\wbn{$\WB_n$ }
\def\WA{\mathop{\it WA}\nolimits}
\def\WB{\mathop{\it WB}\nolimits}
\def\WBC{\mathop{\it WBC}\nolimits}
\def\WD{\mathop{\it WD}\nolimits}
\def\WG{\mathop{\it WG}\nolimits}



\def\zz#1{(z-z')^{#1}}

\openup 1\jot

\pubnumber{DTP 98-49}
\pubdate{July 1998}

\title{{QUANTUM CORRECTIONS TO THE CLASSICAL REFLECTION FACTOR IN $a_2^{(1)}$ TODA FIELD THEORY}
}

\author{
Michael Perkins and Peter Bowcock\thanks{ Email \tt
M.G.Perkins@durham.ac.uk Peter.Bowcock@durham.ac.uk }
\address{Dept. of Mathematical Sciences,
University of Durham,  
Durham, DH1 3LE, U.K.}
}

\abstract{The $O(\beta^2)$ quantum correction to the classical
reflection factor is calculated for one of the integrable boundary
conditions of $a_2^{(1)}$ affine Toda field theory. This is found to
agree
with the conjectured exact reflection factor of the quantum theory. 
We consider the existence of other exact reflection factors consistent
with 
our perturbative answer and examine the question of how duality 
transformations might relate theories with different boundary conditions.}
\maketitle

\section{Introduction}
Despite recent advances, there still remain many unresolved questions regarding 
Toda field theories \cite{EC3} restricted to the half-line. For the full line 
case, exact forms for the $S$-matrix have been postulated and checked to high 
order by perturbation theory; however, for the half-line we have an extra 
quantity $K$ \cite{CHER,SKL}, the factor which encodes reflection off the boundary, which has 
still not been uniquely determined.

Perhaps the most progress has been made for the Sine-Gordon and Sinh-Gordon 
theories. Ghoshal and Zamolodchikov \cite{GZ} calculated the soliton reflection 
factors for the Sine-Gordon model. Ghoshal \cite{GHOSH} used these results together with the reflection bootstrap equations \cite{FK2,FK1,CDR,RYU} 
\be
K_{c}(\theta_{c}) = K_{a}(\theta_{c} +i \overline{\theta}_{ac}^{b}) S_{ab}(2 \theta_c) K_{b} (\theta_{c} - i \overline{\theta}_{bc}^{a}).
\label{rbe}
\ee
to calculate the breather reflection factors. 
Here, indices refer to particle type, and $K$ and $S$ are the reflection and scattering factors respectively. The definitions for rapidities $\theta$ and fusing angles $\theta_{ac}^{b}$ can be found in \cite{BCDS}. 
The reflection factor for the lightest breather can be reasonably identified 
with the particle reflection factor for the Sinh-Gordon model, if the coupling 
constant is suitably analytically continued. Still, not everything is known in 
this case, since it is not known how to relate the two parameters appearing in 
Ghoshal's formula to the two parameter family of boundary conditions allowed by 
integrability in the Sinh-Gordon model, except at special points. One way to 
remove the missing link is to calculate the reflection factor to low order
directly in perturbation theory and compare the result with Ghoshal's 
formula. This was the approach taken by Corrigan in \cite{EC}. 

In this paper we use perturbation theory to gain further 
information on the reflection factors of perhaps the next simplest case; 
that of real-coupling $a_{2}^{(1)}$ affine Toda field theory (ATFT). 
This theory has an important difference with the Sinh-Gordon model 
in that only a finite number
of boundary conditions are compatible with integrability. This property
is  
shared with all the other Toda theories based on affine simply-laced algebras
\cite{BCDR},
and in this sense the $a_2^{(1)}$ theory is more generic 
that the Sinh-Gordon model. Despite the 
fact that only a finite number of possible reflection 
factors exist, less is known about 
this case than the Sine(Sinh)-Gordon models. Again the factorisability of ATFT implies the exact reflection factors must 
satisfy the reflection bootstrap equations (\ref{rbe}), but this does not
determine the solution uniquely. Further information can be obtained by
considering the classical limit, in which we can explicitly calculate the
reflection factor \cite{CDR,PB}, for example by using analytically-continued soliton solutions of the
imaginary coupling theory. The classical solutions satisfy the 
classical limit of (\ref{rbe}), 
\be
K_{c}(\theta_{c}) = K_{a}(\theta_{c} +i \overline{\theta}_{ac}^{b}) K_{b} (\theta_{c} - i \overline{\theta}_{bc}^{a}).
\label{crbe}
\ee
However, knowledge of this limit is still not sufficient to uniquely 
determine the factor in the quantum case.  

Nonetheless, by considering the pole structure and invoking minimality, various
conjectures for the exact quantum reflection factors of $a_2^{(1)}$ were put forward in 
\cite{CDR}. Strong evidence in support of some of these conjectures were 
provided by Gandenberger \cite{GAND}. Using a similar 
approach to Ghoshal, he calculated the breather reflection factors in the 
imaginary coupling theory.
These were then analytically continued to determine the particle reflection
factor in the real coupling theory. In contrast to the situation for the
Sinh-Gordon theory, the discreteness of possible boundary conditions for
$a_2^{(1)}$ ensure that there is no problem in 
identifying the answers with particular boundary conditions. The solutions found in 
\cite{GAND} are consistent with two boundary conditions: Neumann boundary
conditions and another non-trivial boundary condition which in the notation of
\cite{CDR} is referred to as `$+++$'. The aim of the current paper is to test
this solution to the `$+++$' condition by directly calculating the leading 
order correction to the reflection factor for this boundary condition in
perturbation theory. This calculation is in a similar spirit to the 
Sinh-Gordon calculation in \cite{EC}, and to the papers of Kim and Yoon
\cite{KIM,KY} who calculated this term for 
Neumann boundary conditions in a wide variety of theories. Our result is in
agreement with the solution of \cite{CDR,GAND}.

One of the most interesting aspects of studying affine Toda theory on a half-line is to see whether the theory shares the weak-strong coupling duality
symmetry of the bulk theory. Under this symmetry the coupling constant 
undergoes the 
transformation $\beta\to 4 \pi/\beta$. In \cite{KIM,KY}, conjectures for the 
exact reflection factor for Neumann boundary conditions were made based on 
the assumption that the factor should be invariant under this symmetry in
the same way as the bulk S-matrix. Another point of view is that the duality
changes the boundary conditions and that the reflection factor for a theory
with one boundary condition should map to another reflection factor of the 
theory with another (possibly different) boundary condition. Indeed many of 
the conjectures of \cite{CDR} have the latter interpretation, and in particular
the reflection factor for the `$+++$' boundary condition maps to that of the 
Neumann boundary condition. Whilst our perturbative calculation agrees with
this latter solution, 
we point out that there are other `non-minimal' solutions  
with the correct
classical limit and consistent with the bootstrap and our calculation which 
possess very different duality properties. 

In the following section we shall review the properties of affine Toda field theory and establish the notation needed in the rest of the paper. Section three shall deal with the perturbative expansion of $a_2^{(1)}$ with a boundary potential. In section four we shall present our calculations, whose interpretation is dealt with in section five. We take a brief look at the question of duality in section six before presenting our conclusions in section seven.

\section{$a_{2}^{(1)}$ affine Toda field theory}

The Lagrangian for the Toda field theory associated with an affine algebra $\hat{g}$ is given by
\be
{\mathcal{L}}=\frac{1}{2} \partial_\mu \phi \partial^\mu \phi - \frac{m^2}{\beta^2} \sum_{i=0}^{n} n_{i} e^{\beta \alpha^{i} . \phi}.
\ee
Here the $\alpha^{i}$ are the roots of $\hat{g}$, $m$ is the mass scale (which shall, in line with convention, be immediately set to $m=1$) and $\beta$ is the coupling constant of the theory. The $n_i$ are the ``marks'' which all have value 1 in the $a_n^{(1)}$ series of ATFTs which we consider in this paper. 

It is well known that if we restrict the theory to the half-line $x<0$ then there are only a few possible boundary conditions for which the system remains classically integrable \cite{CDR,BCDR,EC2,EC1,CDRS}. These are the Neumann boundary condition, $\partial_{x} \phi = 0$, and, more interestingly, 
\be
\partial_{x} \phi = \sum_{i=0}^{n} A_{i} \alpha^{i} \sqrt{n_{i}} e^{\alpha^{i} . \phi /2}
\label{bcs}
\ee
where the coefficients take values $|A_{i}|=1$. 
Considering the particular case of $a_2^{(1)}$, this is equivalent to a boundary potential of the form
\be
V_{boundary} = -\frac{1}{\beta^2} 2 \sum_{i=0}^{2} A_i (e^{\frac{1}{2} \beta \alpha^{i} . \phi} -1).
\label{boundpot}
\ee
Considering the equations of motion 
\be
\partial_\mu \partial^\mu \phi + \frac{1}{\beta} \sum_{i=0}^{2} \alpha^{i} e^{\beta \alpha^{i} . \phi} = 0
\ee
and using the fact that
\be
\alpha^{0} + \alpha^{1} + \alpha^2 = 0
\ee
we can see that $\phi_{vacuum}=0$ is a possible background solution. However, only in the cases where all the $A_{i}$ are the same sign will this obey the boundary conditions (\ref{bcs}). We will therefore for simplicity only consider the case where all the $A_i=-1$. This corresponds to 
the boundary condition `$+++$' in the notation of \cite{CDR,BCDR} which is based on the overall sign in front of the exponentials appearing in the boundary
potential (\ref{boundpot}). The case where all the $A_{i}=1$ will be neglected here since it is thought to have only marginal stability \cite{FS,MGP} in the classical theory, whilst other boundary conditions require a non-trivial background \cite{EC,MGP} which would make the calculation considerably more difficult. 

The first problem is to find the classical reflection factor. (For simplicity, we shall denote the classical reflection factor throughout by $K$, and the quantum one by $K^q$). We do this as in \cite{PB} by adding in an incoming and outgoing particle (as found by considering analytically-continued multi-soliton solutions of the imaginary-coupling theory) as plane-wave perturbations to the vacuum solution. By solving the equations of motion, with boundary conditions, we find that asymptotically far from the boundary, $\phi = \rho_a e^{-i \omega t} (e^{ikx} + K(k) e^{-ikx})$. Here, $\rho_a$ is the eigenvector of the mass matrix, $M=\sum_{i=0}^{2} \alpha^i \otimes \alpha^i$, associated with the eigenvalue $m_a$. The reflection factor is defined as the relative phase of the two particles. In this case we find that
\be
K \equiv K_{1}=K_{2}=\frac{2ik+m^2}{2ik-m^2}=\frac{2ik+3}{2ik-3}.
\ee
where we now use $m$ to denote the mass of the particles; for $a_2^{(1)}$ the mass $m \equiv m_1 = m_2 =\sqrt{3}$.
It is convenient to write this in terms of the block notation,
\be
(x) \equiv \frac{\sinh(\frac{\theta}{2} + \frac{i \pi x}{2h})}{\sinh(\frac{\theta}{2} - \frac{i \pi x}{2h})},
\ee
where $h$ is the Coxeter number of the Lie algebra; in our case $h=n+1=3$. 
These blocks clearly obey the relations
\be
(-x)=\frac{1}{(x)}
\ee
and
\be
(h+x)=\frac{1}{(h-x)}
\ee
which will be useful when considering the exact bootstrap equation in section V. In this language the classical reflection factor is
\be
K =\frac{(3)}{(1)(2)}.
\ee
It can be readily checked that this obeys the classical reflection bootstrap equation (\ref{crbe}) as required.

With this notation, the bulk $S$-matrix for $a_2^{(1)}$ ATFT \cite{BCDS} also takes a concise form;
\be
S_{11}(\theta)=S_{22}(\theta)=\frac{(2)}{(B)(2-B)} \mbox{,   } S_{12}(\theta)=S_{21}(\theta)=0.
\ee
For the purposes of the reflection bootstrap equation (\ref{rbe}), we require the value of the $S$-matrix at $2\theta$. This can be shown to be;
\be
S_{11}(2\theta)=S_{22}(2\theta)=\frac{(1)(2+\frac{B}{2})(3-\frac{B}{2})}{(\frac{B}{2})(1-\frac{B}{2})(2)(3)}.
\ee

\section{Perturbation theory}

In order to calculate the Feynman Rules for $a_2^{(1)}$ ATFT, we need to expand the bulk and boundary potentials in terms of the coupling constant $\beta$. To do this, it is first useful to write the field $\phi$ in terms of eigenvectors, $\rho_1$ and $\rho_2$, corresponding the two particles. We can do this by looking at the asymptotic form of the $\tau$-function solutions given in \cite{PB,TH}, whence;
\be
\phi = \left( \begin{array}{c} \phi_{1} \\ \phi_{2} \end{array} \right) = \frac{1}{2 \sqrt{2}} \left( \begin{array}{c} (-1+i\sqrt{3}) \Phi - (1+i\sqrt{3}) \overline{\Phi} \\ -(\sqrt{3}+i) \Phi + (-\sqrt{3}+i) \overline{\Phi} \end{array} \right)
\ee
where $\Phi$ and $\overline{\Phi}$ are the fields corresponding to the particle (type 1) and its conjugate (type 2) respectively, and where we have taken a basis for the roots $\alpha^{i}$ to be 
\be 
\alpha^1 = \left( \begin{array}{c} \sqrt{2} \\ 0 \end{array} \right)  \mbox{, }\alpha^2 = \left( \begin{array}{c} -\frac{1}{\sqrt{2}} \\ \sqrt{\frac{3}{2}} \end{array} \right) \mbox{and } \alpha^0 = -\alpha^1 - \alpha^2 = \left( \begin{array}{c} -\frac{1}{\sqrt{2}} \\ -\sqrt{\frac{3}{2}} \end{array} \right) .
\ee
We then find;
\be
V_{bulk}= 3 \Phi \overline{\Phi} + \frac{1}{2} \beta (\Phi^3 + \overline{\Phi}^3) + \frac{3}{4} \beta^2 \Phi^2 \overline{\Phi}^2 +  O(\beta^3)
\ee
and from (\ref{boundpot})
\be
V_{boundary} = \frac{3}{2} \Phi \overline{\Phi} + \frac{1}{8} \beta (\Phi^3 + \overline{\Phi}^3) + \frac{3}{32} \beta^2 \Phi^2 \overline{\Phi}^2 +  O(\beta^3).
\ee

We also need to know the Green's function for the theory. It has been shown \cite{EC} that the Green's function for ATFT on a half-line, in a flat background, takes the form
\be
G(x,t;x',t')=\int \frac{d \omega dk}{(2 \pi)^2} e^{-i\omega (t'-t)} \frac{e^{ik(x'-x)}+K(k)e^{-ik(x+x')}}{\omega^2-k^2-m^2+i \epsilon}
\ee
where $K(k)$ is as before the classical reflection factor. By inspection of the form of the above, it is clear that we can find the $O(\beta^2)$ correction to the classical reflection factor by looking at the coefficient of $e^{-ik(x+x')}$ in the one-loop (i.e. $O(\beta^2)$) contribution to the two-point function.

There are three basic types of diagram which contribute to the two point function in affine Toda field theory \cite{KIM,KIM1}. These are shown in figure 1. The type II diagram is only included for completeness here since it cannot occur in the $a_{n}^{(1)}$ series of ATFTs with a flat background solution $\phi_{vacuum}=0$. This is due to the fact that the loop vertex in this case would require the existence of a $\phi \overline{\phi}^2$ or $\phi^2 \overline{\phi}$ term in the ATFT potential, which does not occur. Note also that we will obtain different contributions to the quantum reflection factor from diagrams where none, one or both of the vertices are located on the boundary $x=0$. There are thus five different diagrams, two of type I and three of type III, which contribute to the two-point function and hence to the quantum correction to the reflection factor. 

\begin{figure}
\begin{center}
\epsfig{figure=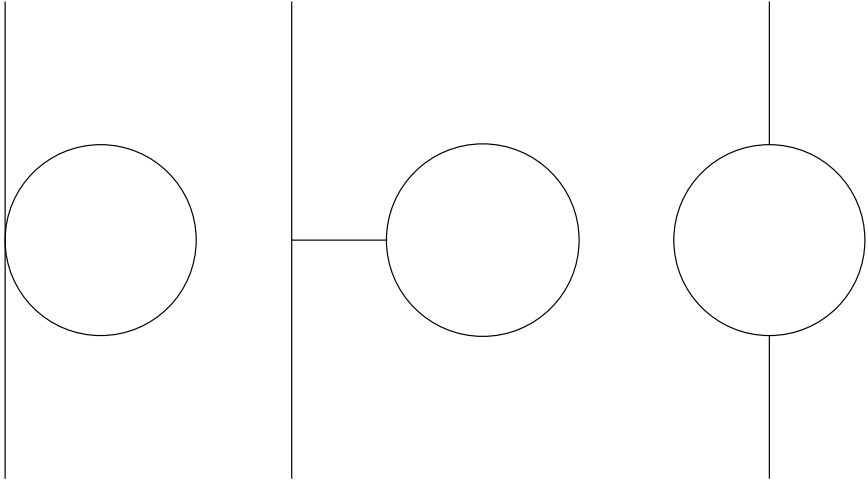}
\caption{Types I, II and III diagrams respectively.
\label{fig:diags}}
\end{center}
\end{figure}

\section{The Calculations}

Let us now proceed to the explicit calculations used to determine the one-loop contribution to the two-point function. We shall consider the two types of Feynman diagram separately in the two following subsections. 

\subsection{Type I diagrams}

There are two possible type I diagrams: one where the vertex is in the bulk region $x<0$, and one where the vertex is situated on the boundary $x=0$ itself. Let us consider the latter first. We require the integral:
\be
I^{I}_{boundary} = -\frac{3}{8} i \beta^{2} \int_{-\infty}^{\infty} dt_{1} G(x,t;0,t_{1}) G(0,t_{1};0,t_{1}) G(0,t_{1};x',t').
\ee
Included here are the vertex and combinatorial factors arising in the four point interaction.

After integrating over $t_{1}$, which generates a delta function, and using this delta function to integrate over $\omega'$, we find
\pagebreak
\begin{eqnarray}
I^{I}_{boundary} &=& -\frac{3}{8} i \beta^2 \int \frac{d\omega dk dk'}{(2 \pi)^3} e^{-i \omega (t'-t)} \frac{i}{p^2 - m^2 + i \epsilon} \frac{i}{p'^2 - m^2 + i \epsilon} \nonumber \\
 & & (1+K) (1+K') e^{-ikx-ik'x'} \left\{ \int \frac{d \omega_1 dk_1}{(2 \pi)^2} \frac{i}{p_1^2 -m^2 +i \epsilon} (1+K_1) \right\}.
\end{eqnarray}
Here we have used the shorthand $p^2 \equiv \omega^2-k^2$, and also the convenient notation $K_{1} \equiv K(k_{1})$ and so on. Notice that the $p_{1}$ integral separates from the others and is divergent due to the $(1+K_{1})$ term. This can be removed by an infinite renormalisation of the boundary potential. In fact, 
\be
1+K_1 = 2 + \frac{2m^2}{2ik_1-m^2} ,
\ee 
and we can perform a minimal subtraction of the divergent part to leave $\frac{2m^2}{2ik_{1} - m^2}$ by adding a suitable counter term.
Then by integrating over $\omega_{1}$ using a contour integral (closing the contour into the lower half-plane), we pick up a pole at
\be
\omega_{1}=\hat{\omega_{1}} = \sqrt{k_{1}^2 + m^2}.
\ee
This leaves us with the task of computing the integral
\be
\int_{-\infty}^{\infty} \frac{dk_{1}}{2 \pi} \frac{1}{2 \hat{\omega_{1}}} \frac{2m^2}{2ik_{1} - m^2}.
\ee
This can be achieved by performing another contour integral which we close in the upper half plane, negotiating the branch cuts which run from $im$ to $i \infty$. Hence we will obtain in general two parts to the integral: integrals along the branch cut, which can be evaluated using the result that 
\be
\int_{m}^{\infty} dy \frac{1}{\sqrt{y^2-m^2}} \frac{1}{y+m \zeta} = \frac{1}{m \sqrt{1-\zeta^2}} (\frac{\pi}{2}-\tan^{-1} \frac{\zeta}{\sqrt{1-\zeta^2}})
\ee
and residues coming from any poles present in the upper half-plane.

Consider the integral we need to compute. There are no poles in the upper half-plane so we only need consider the branch cut contribution. Making the change $k_{1} = iy$ we obtain (noting the factor of 2 obtained by integrating over both sides of the branch cut)
\be
\int_{m}^{\infty} \frac{dy}{\pi} \frac{1}{\hat{\omega_{1}}} \frac{m^2}{-2(y + m(\frac{m}{2}))}
= - \frac{m}{2 \pi \sqrt{1-\frac{m^2}{4}}} (\frac{\pi}{2} - \tan^{-1} (\frac{\frac{m}{2}}{\sqrt{1-\frac{m^2}{4}}})
= -\frac{m}{6} 
\ee 
where in the last step we use the fact that the mass $m=\sqrt{3}$.
Finally, integrating over the remaining $k$ and $k'$ integrals picks up poles at $k=k'= \hat{k} \equiv \sqrt{\omega^2 - m^2}$ and we obtain the result
\begin{eqnarray}
I^{I}_{boundary} &=& \frac{3i}{8} \beta^2 \int \frac{d\omega}{2 \pi} \frac{1}{4 \hat{k}^2} e^{-i\hat{k} (x+x')} e^{-i \omega (t'-t)} (1+\hat{K})^2 \frac{m}{6} \nonumber \\
 &=& \beta^2 \int \frac{d\omega}{2 \pi} \frac{1}{2 \hat{k}} e^{-i\hat{k} (x+x')} e^{-i \omega (t'-t)} \frac{-im \hat{k}}{2(2i \hat{k} - 3)^2}.
\end{eqnarray}

Now let us consider the case where the vertex is located in the bulk section. The contribution this time is
\be
I^{I}_{bulk} = -3 i \beta^{2} \int_{-\infty}^{0} dx_{1} \int_{-\infty}^{\infty} dt_{1} G(x,t;x_{1},t_{1}) G(x_{1},t_{1};x_{1},t_{1}) G(x_{1},t_{1};x',t').
\ee
Again we have a divergency and after an infinite mass renormalisation (in essence performed by simply removing that part of the integrand independent of $x_1$) \cite{EC}, and integration using the delta function in $\omega'$, this becomes
\begin{eqnarray}
-3i \beta^2 \int_{-\infty}^{0} dx_{1} \int \frac{d\omega dk dk'}{(2 \pi)^{3}} e^{-i\omega (t'-t)} \frac{i}{p^2-m^2+i \epsilon} \frac{i}{p'^2-m^2+i \epsilon} (e^{ikx} + Ke^{-ikx}) \nonumber \\
(e^{ik'x'} + K'e^{-ik'x'}) \left\{ \int \frac{d \omega_1 dk_1}{(2 \pi)^2} K_{1} e^{-ix_{1} (k+k'+2k_{1})} \frac{i}{p_{1}^2-m^2+i \epsilon} \right\}.
\label{onebulk}
\end{eqnarray}
To compute this integral, it is useful to note that, if we take $\rho$ to be a small positive quantity, then
\be
\int_{-\infty}^{0} dx e^{ikx+\rho x} = \frac{-i}{k-i \rho}.
\ee
Hence by introducing such a $\rho$ (and taking it to zero at the end of the calculation) we can perform the integral over $x_{1}$. Now consider only the integrations over internal momenta and energies. We need to find
\be
\int \frac{d \omega_{1} dk_{1}}{(2 \pi)^2} \frac{i}{p_{1}^2-m^2+i \epsilon} K_{1} \frac{-i}{-(k+k'+2k_{1})-i \rho}.
\ee
As before, integration over $\omega_{1}$ picks up the pole at $\hat{\omega_{1}}$ and we can complete the integration over $k_{1}$ by decomposing into partial fractions and using the same technique as before. We must then perform the $k$ and $k'$ integrations. This is done as before by integrating along a complex contour closed in such a direction that the exponential factors of each term decay to zero on the complex part of the contour. The only poles we pick up are simply $\pm \hat{k}$ since it can be checked that all other poles have a finite imaginary part and hence the exponentials decay to zero if we take the limits $x, x' \rightarrow -\infty$.

It turns out to be simpler if, knowing this fact, we simplify the integrand of the $k_{1}$ integral by substituting in the poles of the $k$ and $k'$ integration first, i.e. if we define the previous integrand to be $I$, then the new integrand is
\be
 \frac{1}{4 \hat{k}^2} e^{-i\hat{k} (x+x')} (I|_{k=-\hat{k}, k'=-\hat{k}} + \hat{K} I|_{k=-\hat{k}, k'=\hat{k}} + \hat{K} I|_{k=\hat{k}, k'=-\hat{k}} + \hat{K}^2 I|_{k=\hat{k}, k'=\hat{k}}).
\ee
The calculation is made simpler still by noting that we need only consider that part of the integrand which is even in $k_{1}$, and by making the substitutions $\omega = m \cosh (\theta)$ and $e^{\theta} = y$. The answer then consists of a sum of contributions from integrals along branch cuts, residues arising from poles with finite imaginary part, and residues from poles which are only infinitesimally shifted from the real axis. In the first two cases, the infinitesimals present in the integrand are insignificant to the calculation. We can therefore reduce the calculation to a manageable size by setting all infinitesimals to zero, and simplifying the integrand, before calculation of these two contributions. However, in the case of poles with infinitesimal imaginary part, no such simplification can be made. 

Hence the integration over the branch cuts and poles with finite imaginary part becomes
\be
\int \frac{d \omega}{2 \pi} \int_{-\infty}^{\infty} \frac{dk_{1}}{2 \pi} \frac{1}{2 \hat{k}} e^{-i \omega (t'-t)} e^{-i \hat{k} (x+x')} \frac{54 i (y^2-1)}{(2 i \hat{k} -3)^2 (2k_{1} +3i) (2k_{1}-3i) \sqrt{3} y} .
\ee
We integrate using the same method as before, but this time we encounter a pole at $k_{1}=\frac{3}{2} i$ and hence must include the contribution arising from its residue. Summing the contributions gives
\be 
\int \frac{d \omega}{2 \pi} \frac{1}{2 \hat{k}} e^{-i \omega (t'-t)} e^{-i \hat{k} (x+x')} \frac{ i(y^2-1)}{(2 i \hat{k} -3)^2 y} .
\ee
The integral over the poles which have been shifted off the real axis by an infinitesimal amount involves considering the residues of poles in the upper half-plane of
\begin{eqnarray}
\frac{9 \sqrt{3} (3-8k_1^2y^2+3y^2-12ik_1y^2+3y^4)(2k_1-3i)(y^2-1)}{2 y^2 (3y^2-3+2\sqrt{3}k_1y+i \sqrt{3} \rho y)(3y^2-3-2\sqrt{3}k_1y-i \sqrt{3} \rho y)(2k_1+i \rho)(2k_1+3i)} \nonumber \\
+ (k_1 \rightarrow -k_1)
\end{eqnarray}
which yield a contribution
\be
-\int \frac{d \omega}{2\pi} \frac{1}{2 \hat{k}} \frac{3i}{4} \frac{(y^2+y+1)(y^2-y+1)(y-1)}{(2i\hat{k} -3)^2 y(y^2+1)(y+1)}.
\ee
The reason for leaving the factors $\frac{1}{2 \hat{k}}$ and $\frac{1}{(2i\hat{k}-3)^2}$ in terms of $\hat{k}$ rather than $y$ will become clear later.

\subsection{Type III diagrams}

In the case of the type III diagrams, there are three possible configurations - we can have none, one, or both of the vertices located on the boundary. The simplest of these is the last --- the boundary-boundary case;
\be
I^{III}_{bndry-bndry} = -\frac{9}{32} \beta^{2} \int_{-\infty}^{\infty} dt_{1} dt_{2} G(x,t;0,t_{1}) G(0,t_{1};0,t_{2}) G(0,t_{1};0,t_{2}) G(0,t_{2};x',t').
\ee
Let us consider this case in detail since it is instructive for performing the later integrals, which are more tedious.
Putting in the form of $G$, we obtain
\begin{eqnarray}
-\frac{9}{32} \beta^2 \int_{-\infty}^{\infty} dt_{1} dt_{2} \int \frac{dp^2 dp'^2 dp_{1}^2 dp_{2}^2}{(2 \pi)^8} e^{-i\omega (t_{1}-t)-i\omega_{1}(t_{2}-t_{1})-i\omega_{2}(t_{2}-t_{1})-i\omega' (t'-t_{2})} \nonumber \\
\frac{i}{p^2-m^2+i\epsilon} \frac{i}{p'^2-m^2+i\epsilon} \frac{i}{p_{1}^2-m^2+i\epsilon} \frac{i}{p_{2}^2-m^2+i\epsilon} \nonumber \\
e^{-ikx-ik'x'} (1+K)(1+K')(1+K_{1})(1+K_{2}) .
\end{eqnarray} 
As before, integrating over the $t_{1}$ and $t_{2}$ gives us delta functions which enable integration over $\omega'$ and $\omega_{2}$. However, this means that we must not only set $\omega'=\omega$ as before but we also have $\omega_{2}=\omega-\omega_{1}$. Substituting in the form of $K_1$ and $K_2$, we find that we must perform the integral
\begin{eqnarray}
-\frac{9}{32} \beta^2 \int \frac{d \omega dk dk'}{(2 \pi)^3} \frac{i}{p^2-m^2+i\epsilon} \frac{i}{p'^2-m^2+i\epsilon} e^{i \omega (t-t')} e^{-ikx-ik'x'} (1+K) (1+K') \nonumber \\
\left\{ \int \frac{d\omega_{1} dk_{1} dk_{2}}{(2 \pi)^3} \frac{i}{\omega_{1}^2 - k_{1}^2 -m^2 +i\epsilon} \frac{i}{(\omega-\omega_{1})^2 - k_{2}^2 -m^2 +i\epsilon} \frac{4ik_{1}}{2ik_{1}-m^2} \frac{4ik_{2}}{2ik_{2}-m^2} \right\}.
\end{eqnarray}
Consider the integral in the second bracket. Integrating over $\omega_{1}$ (again closing the contour downwards) we pick up two poles, at $\sqrt{k_{1}^2+m^2}$ and $\sqrt{k_{2}^2+m^2} + \omega$. Hence we obtain
\begin{eqnarray}
\int \frac{dk_{1} dk_{2}}{(2\pi)^2} \frac{-16k_{1} k_{2}}{(2ik_{1}-m^2)(2ik_{2}-m^2)} \left( \frac{1}{2 \sqrt{k_{1}^2+m^2}} \frac{i}{(\omega-\sqrt{k_{1}^2+m^2})^2-k_{2}^2-m^2+i\epsilon} + \right. \nonumber \\
\left. \frac{1}{2 \sqrt{k_{2}^2+m^2}} \frac{i}{(\omega + \sqrt{k_{2}^2+m^2})^2 - k_{1}^2 - m^2 +i\epsilon} \right) 
\end{eqnarray}
Notice however that we can simply exchange the indices $1$ and $2$ on the second term giving us;
\be
\int \frac{dk_{1} dk_{2}}{(2\pi)^2} \frac{1}{2 \hat{\omega_{1}}} \frac{i}{(\omega-\hat{\omega_{1}})^2-k_{2}^2-m^2+i\epsilon} \frac{-16k_{1} k_{2}}{(2ik_{1}-m^2)(2ik_{2}-m^2)} + (\omega \rightarrow -\omega)
\ee
where $\hat{\omega_{1}} \equiv \sqrt{k_{1}^2+m^2}$.

Now integrating this over $k_{2}$ (closed in the upper half-plane) gives a residue due to the pole at $\hat{k_{2}} \equiv \sqrt{(\omega-\hat{\omega_{1}})^2-m^2}$. So we are left with the integral
\be
\int \frac{dk_{1}}{2\pi} \frac{1}{\hat{\omega_{1}}} \frac{-4k_{1}}{(2ik_{1}-m^2)(2i\hat{k_{2}}-m^2)} + (\omega \rightarrow -\omega).
\ee
This can be easily decomposed into two pieces: that which contains odd powers of $\hat{k_{2}}$ and that which does not. Moreover, if we throw away all terms which are odd in $k_{1}$ (we are integrating over all $k_1$), we are left with
\be
\int \frac{dk_{1}}{2\pi} \frac{1}{\hat{\omega_{1}}} \left( \frac{-8im^2 k_{1}^2}{(4k_{1}^2+m^4)(4\hat{k_{2}}^2 +m^4)} + \frac{16 k_{1}^2 \hat{k_{2}}}{(4k_{1}^2+m^4)(4\hat{k_{2}}^2 +m^4)} \right) + (\omega \rightarrow -\omega).
\ee
The first term in the above can be handled as before. The second term, however, is difficult to deal with since it generates elliptic integrals.  

At this stage it is worth doing a little analysis of the properties we expect of the results. Unitarity implies that the quantum reflection factor, $K^q$ is a pure phase, i.e. of the form $e^{i \chi}$. Suppose that the classical reflection factor is $K=e^{i \chi_{0}}$. Then the quantum reflection factor, to order $\beta^2$, is given by $e^{i(\chi_{0} + \beta^2 \chi_{1})}$ ($\chi_0$ and $\chi_1$ some functions of $k$). Expanding this, we obtain
\be
K^q=e^{i(\chi_{0} + \beta^2 \chi_{1})} = e^{i\chi_{0}} (1+ i \beta^2 \chi_{1} + O(\beta^4)) = K + iK \chi_{1} \beta^2 + O(\beta^4).
\ee

Hence we are looking for a $\beta^2$ correction which is what we shall term ``completely imaginary with respect to the phase of the classical reflection factor $K$'', i.e. its argument is $\arg K + \frac{\pi}{2}$. Notice that the phase of $(1+K)^2$ is the same as that of $K$, and this is exactly the prefactor we obtain from the $k$ and $k'$ integrals. So there is good physical justification for assuming that all the completely real parts (which in every case are the ``elliptic'' parts) of the integrals will eventually, though perhaps only after summation of terms from all diagrams, vanish. This assumption shall make the job of calculating these integrals significantly simpler.

Hence ignoring the real part of this integral, and using the expression for $\hat{k_{2}}$, we obtain
\be
\int \frac{dk_{1}}{2\pi} \frac{1}{\hat{\omega_{1}}} \frac{-8im^2 k_{1}^2}{(4k_{1}^2+m^4)(4\hat{k_{2}}^2 +m^4)} + (\omega \rightarrow -\omega)
\ee
\be
= \int \frac{dk_{1}}{2\pi} \frac{1}{\hat{\omega_{1}}} \frac{-16i k_{1}^2 m^2 (4\omega^2 + 4k_{1}^2 +m^4)}{(4k_{1}^2+m^4)((4\omega^2+4k_{1}^2+m^4)^2-64\omega^2 (k_{1}^2+m^2))}.
\ee
This integral can be computed as before and is found to contribute undesirable $\tanh^{-1}$ terms to the $O(\beta^2)$ correction. Fortunately these are found to exactly cancel with terms from the other two type III diagrams.

In a similar way to before, we find the necessary $k_{1}$ integrals for the other type III contributions to be
\be
I_{bulk-bndry}^{III} = -\frac{9}{4} \beta^2 \int \frac{dk_{1}}{2\pi} \frac{1}{\hat{\omega_{1}}} \frac{4ik_{1}}{(k+k_{1}+\hat{k_{2}})(2ik_{1}-m^2)(2i\hat{k_{2}}-m^2)} + (\omega \rightarrow -\omega)
\ee
and
\begin{eqnarray}
I_{bulk-bulk}^{III} &=&-\frac{9}{2} \beta^2 \int \frac{dk_{1}}{2\pi} \left\{ \frac{1}{4 \hat{\omega_{1}} \hat{k_{2}}} \frac{1}{k_{1}+\hat{k_{2}}+k+i \rho} \left( \frac{1}{k'-k_{1}-\hat{k_{2}}+i\rho} \right. \right. \nonumber \\
&+& \left. K_{1} \frac{1}{k_{1}-\hat{k_{2}} +k'+i\rho} + \hat{K_{2}} \frac{1}{\hat{k_{2}}-k_{1}+k'+i \rho} + K_{1} \hat{K_{2}} \frac{1}{k_{1}+\hat{k_{2}}+k'+i\rho} \right) \nonumber \\
&+& \frac{1}{2 \hat{\omega_{1}}} \frac{1}{(\omega-\hat{\omega_{1}})^2-(k'-k_{1})^2-m^2} \frac{1}{k+k'+i\rho} \nonumber \\
&+& \left. \frac{1}{2 \hat{\omega_{1}}} K_{1} \frac{1}{(\omega-\hat{\omega_{1}})^2-(k_{1}+k')^2-m^2} \frac{1}{2k_{1}+k+k'+i \rho} \right\} \nonumber \\
&+& (\omega \rightarrow -\omega)
\label{bulkbulk}
\end{eqnarray}
before separation of the real and imaginary parts relative to the phase. The ubiquitous $k$ and $k'$ integrations take the same form as in (\ref{onebulk}).

Let us consider the fifth term of the integral (\ref{bulkbulk}), whose analysis is quite subtle. It is clear that this term contains a double pole in $k$ and hence we need to perform some finite mass renormalisation \cite{BS,DDV}. This is equivalent to adding a term of the form $a \Phi \overline{\Phi}$, where $a$ is a constant, to the original Lagrangian in order to cancel out this double pole. The contribution arising from such a renormalisation term is
\be
a \int_{-\infty}^{\infty} dt_1 \int_{-\infty}^{0} dx_1 G(x,t;x_1,t_1) G(x_1,t_1;x',t') 
\ee
which can be manipulated to give
\begin{eqnarray}
a \int \frac{d\omega dk dk'}{(2 \pi)^3} e^{-i \omega (t'-t)} \frac{i}{\omega^2-k^2-m^2+i \epsilon} \frac{i}{\omega^2 - k'^2-m^2+i \epsilon} \frac{i}{k+k'+i \rho} \nonumber \\
 (e^{ikx} +K e^{-ikx})(e^{ik'x'}+K'e^{-ik'x'}).
\end{eqnarray}
Now look at the double pole term in our integral. This has broadly the above form, but instead of the constant $a$, we have a function, $\int dk_1 f(k_1,k',\omega)$ say, of $\omega$ and $k'$. Looking at its precise form in (\ref{bulkbulk}), and discarding all odd terms in $k_1$, we can see that $f$ is also even in $k'$. Hence by substituting $\omega^2=\hat{k}^2+m^2$ into this function and Taylor expanding the result around $k'^2=\hat{k}^2$ we obtain
\be
f(k_1,k',\omega) = f_0(k_1,\hat{k}) + f_1(k_1,\hat{k}) (k'^2-\hat{k}^2) + O((k'^2-\hat{k}^2)^2).
\label{wfrenex}
\ee
Integrating over the $k_1$, the first term in the above expansion gives a constant independent of $\hat{k}$. This can then be cancelled off by adding in an equivalent finite mass renormalisation term. In fact, we obtain
\be
F_0 = \int \frac{dk_1}{2 \pi} f_0(k_1,\hat{k}) = \frac{-\sqrt{3}}{12} \beta^2
\ee
which is exactly the same mass renormalisation as that obtained in the full-line case \cite{BS}, as we would expect.

The second term in the expansion (\ref{wfrenex}) is more interesting. This gives us
\be
\int \frac{d \omega dk dk'}{(2\pi)^3} i \frac{i}{\omega^2-k^2-m^2+i\epsilon} \frac{i}{k+k'+i \rho} (e^{ikx} +K e^{-ikx})(e^{ik'x'}+K'e^{-ik'x'}) F_1(\hat{k}) \ee
where $F_1(\hat{k})=\int dk_1 f_1(k_1,\hat{k})$. By integrating over $k'$ first and then $k$ we obtain
\be
\int \frac{d \omega}{2 \pi} \frac{1}{2 \hat{k}} i F_1(\hat{k}) (e^{i\hat{k} (x'-x)} + \hat{K} e^{-i\hat{k} (x+x')}).
\label{wavfnterm}
\ee
The value of $F_1(\hat{k})$ can be calculated and it is found that, as must be the case, it is simply a number, not dependent on $\hat{k}$. The value is
\be
F_1 = (-\frac{1}{36 \sqrt{3}} - \frac{1}{12 \pi}) \beta^2.
\ee

Notice that this term has changed our coefficient of $e^{i \hat{k} (x'-x)}$. We must perform a wavefunction renormalisation in order to return the coefficient of this term to unity, or in other words to cancel out the first term in (\ref{wavfnterm}). This can be done by rescaling $\Phi$ and $\overline{\Phi}$ by 
\be
\Phi \overline{\Phi} \rightarrow (1-F_1) \Phi \overline{\Phi}.
\ee
This rescales the propagator by the same amount, cancelling out the entirety of (\ref{wavfnterm}). It can thus be seen that renormalisation allows us to completely discard the fifth term of (\ref{bulkbulk}).
 
The calculations from here are tedious and it is worth noting that as before, considerable simplifications can be made by using the values of $k$ and $k'$ given by their poles and simplifying the integrand. By adding all type III integrands, discarding the elliptic parts (which it can be checked are always completely real w.r.t. the phase), and considering only those parts even in $k_{1}$, we finally obtain 
\be 
\beta^2 \int \frac{d\omega}{2\pi} e^{-i\omega (t'-t)} e^{-i\hat{k} (x+x')} \frac{1}{2 \hat{k}} \frac{-i(2y^2+y+2)(y^2+3y+1)(y-1)}{(2i\hat{k}-3)^2 4y(y^2+1)(y+1)}.
\ee

Adding this result to those of the type I integrals yields a total contribution
\be 
\beta^2 \int \frac{d\omega}{2\pi} e^{-i\omega (t'-t)} e^{-i\hat{k} (x+x')} \frac{1}{2 \hat{k}} \frac{-i(4y^2+5y+4)(y-1)}{(2i\hat{k}-3)^2 4y(y+1)}
\ee
and hence the na\"{\i}ve $\beta^2$ correction to the reflection factor is
\be
K^{\beta^2} = \frac{-i(4y^2+5y+4)(y-1)}{(2i\hat{k}-3)^2 4y(y+1)} \beta^2.
\label{perans}
\ee
In the next section we shall show  that it is necessary to carry out a further finite renormalisation in order to make sense of this result.

\section{Interpretation}

We now wish to test whether our answer obeys the reflection bootstrap equation.
Consider the reflection bootstrap equation at $O(\beta^2)$, which reads for $a_{2}^{(1)}$ ATFT (taking $K^q = K + K^{\beta^2} + O(\beta^4)$):
\begin{eqnarray}
K_2^{\beta^2}(\theta) = K_1^{\beta^2}(\theta +i \frac{\pi}{3}) K_1 (\theta - i \frac{\pi}{3}) &+& K_1(\theta +i \frac{\pi}{3}) K_1^{\beta^2} (\theta - i \frac{\pi}{3}) \nonumber \\ 
&+& K_1(\theta +i \frac{\pi}{3}) S_{11}^{\beta^2}(2 \theta) K_1(\theta - i \frac{\pi}{3}) .
\label{qboot}
\end{eqnarray}
Here, subscripts denote particle type.

Let us assume that the quantum reflection factors for particles 1 and 2 are equal (and hence that $K_1^{\beta^2}=K_2^{\beta^2}$), as would seem sensible since they have equal classical limits and identical calculations for the $O(\beta^2)$ correction. Then it is found that the $O(\beta^2)$ correction (\ref{perans}) calculated above does not obey the bootstrap equation (\ref{qboot}). This would appear to be a severe problem. However, let us consider adding a finite counter term of the form $\alpha \beta^2 \Phi \overline{\Phi}$, where $\alpha$ is some coefficient, into the boundary potential. It can quickly be shown that this yields a contribution 
\be
- \alpha \beta^2 \int \frac{d \omega}{2 \pi}  e^{-i \omega (t-t')} e^{-i\hat{k} (x+x')} \frac{1}{2 \hat{k}} 4 \sqrt{3} \frac{y^2-1}{y (2i\hat{k}-3)^2}.
\ee
Notice that this does not change the {\em form} of the propagator, but merely adds another contribution to the $O(\beta^2)$ correction to $K$. Hence we have a freedom to add in such a counter term and change our $O(\beta^2)$ result by the according amount;
\be
- 4 \sqrt{3} \alpha \frac{y^2-1}{y (2i\hat{k}-3)^2} \beta^2.
\label{ambig}
\ee
Suppose that we call our result (\ref{perans}) of the perturbative calculation $f$, and the correction (\ref{ambig}) above $\alpha g$. Now suppose that $f+\alpha g$ obeys the bootstrap. Then we can find $\alpha$ by rearranging (\ref{qboot}), i.e.
\be
\alpha = \frac{f(\theta)-K(\theta +i \frac{\pi}{3})f(\theta - i \frac{\pi}{3}) - K(\theta -i \frac{\pi}{3})f(\theta + i \frac{\pi}{3}) - K(\theta +i \frac{\pi}{3}) S_{11}^{\beta^2}(2 \theta) K(\theta - i \frac{\pi}{3}) }{K(\theta +i \frac{\pi}{3})g(\theta - i \frac{\pi}{3}) - K(\theta -i \frac{\pi}{3})g(\theta + i \frac{\pi}{3}) - g(\theta)}.
\ee
If this gives $\alpha$ as a number (as opposed to a function of $y$) then we know that we can satisfy the reflection bootstrap equation in this way. In fact, using our $\beta^2$ correction (\ref{perans}), we find $\alpha=\frac{-1}{16 \sqrt{3}}$, which gives our total $O(\beta^2)$ correction (i.e. $f+\alpha g$) as
\be
\frac{-3(y^3-1)}{4y(y+1)(2i\hat{k}-3)^2} \beta^2.
\label{correc}
\ee

This correction satisfies the reflection bootstrap equations (\ref{qboot}) to second order in $\beta$, and is the main result of this paper.

The idea that a finite renormalisation of the boundary conditions $A_i$ is necessary to retain integrability of $a_n^{(1)}$ ATFT is not new. Penati et al \cite{PZ,PZ1} have discussed the renormalisation of $a_2^{(1)}$ and their results agree qualitatively with those presented here.

It is now interesting to consider the possible candidates for the exact form of the reflection factor. A possible exact reflection factor - postulated in \cite{CDR} for the boundary condition being considered here - which has the correct classical limit, obeys the reflection bootstrap equation (\ref{rbe}), and appears to be minimal, is 
\be
K^q=\frac{(3-\frac{B}{2})}{(1-\frac{B}{2})(2)}
\label{neu}
\ee
where
\be
B=\frac{\frac{\beta^2}{2 \pi}}{1+\frac{\beta^2}{4 \pi}}
\ee
Further evidence that this is indeed the correct reflection factor was provided by Gandenberger \cite{GAND} whose method based on analytically continued breather reflection matrices in the imaginary coupling theory produced the same answer. Expanding (\ref{neu}) in powers of $\beta$ reproduces the $O(\beta^2)$ correction (\ref{correc}) found above. Thus our perturbative answer is in agreement with the exact reflection factor (\ref{neu}) found by other methods and is a highly non-trivial check of these results.

There are of course other exact reflection factors which obey the bootstrap, have the correct classical limit, and the same $O(\beta^2)$ quantum correction. These can be obtained by multiplying the minimal $K$ matrix by suitable factors. This situation is nothing new: similar ambiguities occur for the bulk $S$-matrix. In this case, such ambiguities have been removed by a careful consideration of the poles on the physical strip that extra factors would introduce. Such an analysis applied to the reflection factors is beyond the scope of the present paper. Instead, we will analyse the possible forms of such ambiguities and discuss their duality properties in the next section.

We shall consider additional factors of the form
\be
F_{C,D}=\frac{(1-C)(1+C)(2-C)(2+C)}{(1-D)(1+D)(2-D)(2+D)} 
\ee
where $C$ and $D$ are two functions of $\beta$ which tend to the same limit as $\beta \rightarrow 0$. This has classical limit 
\be
\lim_{\beta \rightarrow 0} F_{C,D}=1
\ee
and satisfies
\be
F_{C,D}(\theta)= F_{C,D}(\theta - i \frac{\pi}{3})F_{C,D}(\theta + i \frac{\pi}{3})
\ee
so that additional solutions to the bootstrap equation can be generated by multiplying any previous solution by this factor.
By choosing the functions $C$ and $D$ carefully we can ensure that the $O(\beta)$ term vanishes (needed in order to fit with perturbation theory which predicts that there be no $O(\beta)$ term). The necessary condition is
\be  
\frac{dC}{d\beta}(0)=\frac{dD}{d\beta}(0).
\label{ders}
\ee
We can also make the $O(\beta^2)$ term disappear. For the case where $C$ and $D \rightarrow 0$ as $\beta \rightarrow 0$, (\ref{ders}) is a sufficient condition.

For simplicity, let us consider only cases where $C$ and $D$ take the form $\frac{n}{2} \pm \frac{B}{2}$ where $n$ is an integer. Whilst there are of course many other possible forms for $C$ and $D$, it is blocks of the type $(\frac{n}{2} \pm \frac{B}{2})$ which are most commonly postulated to make up the exact reflection factors. For these cases we have $\frac{dC}{d\beta}(0)=\frac{dD}{d\beta}(0)=0$ and find that there are four fundamental factors from which all others can be generated. These are tabulated in Table 1. 
\vspace{0.1in}
\footnotesize

\begin{tabular}{|c|c|c|c|} \hline
\multicolumn{4}{|c|}{Table 1: Factors with which to generate solutions to the reflection bootstrap equations.} \\ \hline
 & & & Classical limit \\ 
Factor & Form & $O(\beta^2)$ term & of dual \\ \hline
$F_1=F(\frac{B}{2},0)$ & $\frac{(1-\frac{B}{2})(1+\frac{B}{2})(2-\frac{B}{2})(2+\frac{B}{2})}{(1)^2 (2)^2}$ & 0 & $\frac{(3)}{(1)(2)}$ \\ \hline
$F_2=\frac{F(\frac{B}{2},0)}{F(2+\frac{B}{2},2)}$ & $\frac{(\frac{B}{2})(1-\frac{B}{2})(1+\frac{B}{2})^2(2-\frac{B}{2})^2(2+\frac{B}{2})(3-\frac{B}{2})}{(1)^3 (2)^3 (3)}$ & $\frac{iy(y^4+1)}{2(y^6-1)}$ & 1 \\ \hline
$F_3=\frac{F(\frac{5}{2}+\frac{B}{2},\frac{5}{2})^2}{F(\frac{3}{2}+\frac{B}{2},\frac{3}{2})}$ & $\frac{(\frac{1}{2}-\frac{B}{2})(\frac{1}{2}+\frac{B}{2})(\frac{3}{2}-\frac{B}{2})^2(\frac{3}{2}+\frac{B}{2})^2 (\frac{5}{2}-\frac{B}{2})(\frac{5}{2}+\frac{B}{2})}{(\frac{1}{2})^2 (\frac{3}{2})^4 (\frac{5}{2})^2}$ & 0 & $(\frac{1}{2})(\frac{3}{2})^2 (\frac{5}{2})$ \\ \hline
$F_4=F(\frac{5}{2}+\frac{B}{2},\frac{5}{2})$ & $\frac{(\frac{1}{2})(\frac{3}{2})^2 (\frac{5}{2})}{(\frac{1}{2}+\frac{B}{2})(\frac{3}{2}-\frac{B}{2})(\frac{3}{2}+\frac{B}{2})(\frac{5}{2}-\frac{B}{2})}$ & $\frac{-iy(y^2-1) \sqrt{3}}{6(y^4-y^2+1)}$ & 1 \\ \hline
\end{tabular} 
\vspace{0.1in}
\normalsize
\openup 1\jot

We can multiply our previous solution (\ref{neu}) by any prefactor consisting of powers of $F_1$, $F_2$, $F_3$ or $F_4$, giving us a new solution to the reflection bootstrap equations (\ref{rbe}). Moreover, if the prefactor consists only of powers of $F_1$ and $F_3$ then this new solution will not be distinguishable from our previous solution by its $O(\beta^2)$ correction. The differences between these solutions leads us naturally into the question of duality.

\section{Duality}

The Affine Toda theories possess a remarkable non-perturbative weak-strong coupling duality. The bulk $S$-matrix is left invariant under the transformation $\beta \rightarrow 4 \pi/\beta$, or in terms of $B$, $B \rightarrow 2-B$. It is plausible that the theory with a boundary shares this symmetry. One way that this could be realised is that $K$ itself is invariant under $\beta \rightarrow 4 \pi/\beta$ and this indeed was advocated by Kim \cite{KIM}. It is also possible however that the symmetry is realised in a more subtle manner, and that under duality a theory with one boundary condition is mapped to a theory with a second boundary condition. 

Let us suppose that the correct exact reflection factor is that given by (\ref{neu}). The dual of the minimal reflection factor is
\be
\frac{(2+\frac{B}{2})}{(\frac{B}{2})(2)}.
\label{neumann}
\ee
First, note that this obeys the reflection bootstrap equation, as it clearly must as the scattering matrix $S$ is self-dual. We can now ask whether this corresponds to any known reflection factor. Clearly since the classical limit is unity, this cannot correspond to the reflection factor associated with any of the boundary conditions (\ref{bcs}), but it could correspond to the reflection factor associated with the Neumann boundary condition. Kim \cite{KIM} has performed an analogous perturbative calculation to the one presented here for the Neumann boundary condition and has determined the $O(\beta^2)$ correction to the classical reflection factor. Although he concentrates on the assumption that the reflection factor must be self-dual, and hence decides on a different exact form from (\ref{neumann}) above, it is found that his reflection factor agrees with this to $O(\beta^2)$. Hence our results are consistent with the `$+++$' and Neumann boundary conditions being related by a duality transformation.

How would this conclusion be changed if we were to consider a non-minimal reflection factor? Suppose we multiply the minimal solution (\ref{neu}) by $F_1$, $F_1^{-1}$ and $F_1 F_3$, leading to the reflection factors
\be
\frac{(1+\frac{B}{2})(2-\frac{B}{2})(2+\frac{B}{2})(3-\frac{B}{2})}{(1)^2 (2)^3},
\label{sd}
\ee 
\be
\frac{(1)^2 (2) (3-\frac{B}{2})}{(1-\frac{B}{2})^2 (1+\frac{B}{2})(2-\frac{B}{2})(2+\frac{B}{2})},
\label{othbc}
\ee
and
\be
\frac{(\frac{1}{2})^2 (\frac{3}{2})^4 (\frac{5}{2})^2 (1+\frac{B}{2})(2-\frac{B}{2})(2+\frac{B}{2})(3-\frac{B}{2})}{(\frac{1}{2}-\frac{B}{2})(\frac{1}{2}+\frac{B}{2})(\frac{3}{2}-\frac{B}{2})^2 (\frac{3}{2}+\frac{B}{2})^2 (\frac{5}{2}-\frac{B}{2})(\frac{5}{2}+\frac{B}{2}) (1)^2 (2)^3}.
\label{mix}
\ee
respectively. These cannot be distinguished from each other and from (\ref{neu}) by the $O(\beta^2)$ term alone.

Look first at (\ref{sd}). It is easy to see that this is self-dual: it transforms into itself under $\beta \rightarrow 4\pi/\beta$. On the other hand, (\ref{othbc}) transforms into
\be
\frac{(1)^2 (2) (2+\frac{B}{2})}{(\frac{B}{2})^2 (1+\frac{B}{2}) (2-\frac{B}{2})(3-\frac{B}{2})}
\ee 
which in the classical limit becomes
\be
(1)(2)(3).
\ee 
This is the classical reflection factor associated with the boundary condition where all the $A_i=1$, i.e. `$---$'.
Finally, taking the classical limit of the dual to (\ref{mix}) we obtain
\be
\frac{(\frac{1}{2})(\frac{3}{2})^2 (\frac{5}{2})}{(1)(2)(3)}
\ee 
which is the classical reflection factor associated with the two boundary conditions where not all the $A_i$ are the same sign \cite{CDR}. So even given the $O(\beta^2)$ correction to the classical reflection factor, we still cannot determine the duality properties of the exact quantum reflection factors.

It is possible to suppose that the boundary conditions are related in pairs by a duality transformation. On the other hand, it may be that the reflection factors are self-dual, or have no duality symmetries. Indeed, even if we knew all the $O(\beta^2)$ corrections for all different boundary conditions, and could postulate consistent reflection factors which related these in pairs under duality transformations, it would still always be possible to generate self-dual reflection factors by use of $F_1$ and $F_3$ which would be equally valid. From another perspective, however, it seems that if we knew both the $O(\beta^2)$ correction and the mappings between the various boundary conditions under duality transformations, we should be able to pin down the exact quantum reflection factor of the theory. 

What we have not considered is the use of more general $C$ and $D$ functions. By multiplying the minimal reflection factor by some $F_{C,D}$ with appropriate choices of $C$ and $D$ it may be possible to create more factors which fit with the perturbative answer. However we shall leave this possibility for future analysis.

\section{Conclusions}

The perturbative result above has been useful in that it has provided further strong evidence in support of the exact quantum reflection factor (\ref{neu}). This (and indeed the result found by Kim for the Neumann boundary condition \cite{KIM}) is in agreement with the hypothesis that the `$+++$' and Neumann boundary conditions are related by a duality transformation. However, it should not be ignored that there exist other exact reflection factors, albeit non-minimal, which also are in agreement with our perturbative result, and have vastly different duality properties. 

In order to place further bounds on the form of the exact reflection factor one could consider extending the perturbative calculations to $O(\beta^4)$ or higher. It is, however, expected that this could involve prohibitively laborious computations. Alternatively, it may also be possible to restrict possibilities for the reflection factors by considering the associated boundary bound states \cite{FK} and pole structures present.

\section{Acknowledgements}

The authors would like to thank E. Corrigan, P. Dorey and G. Gandenberger for enlightening discussions on this topic. MGP also acknowledges the support of the UK Engineering and Physical Sciences Research Council.

\end{document}